# Conversion of Self-Assembled Monolayers into Nanocrystalline Graphene: Structure and Electric Transport

Andrey Turchanin[1]*, Dirk Weber[2], Matthias Büenfeld[1], Christian Kisielowski[3], Mikhail V. Fistul[4], Konstantin B. Efetov[4], Thomas Weimann[2], Rainer Stosch[2], Joachim Mayer[5], and Armin Gölzhäuser[1]

[1]*Physics of Supramolecular Systems and Surfaces, University of Bielefeld, 33615 Bielefeld, Germany*
[2]*Physikalisch-Technische Bundesanstalt, 38116 Braunschweig, Germany*
[3]*National Center for Electron Microscopy, Berkeley, CA 94720, USA*
[4]*Theoretical Physics III, Ruhr-Universität Bochum, 44801 Bochum, Germany*
[5]*Central Facility for Electron Microscopy, RWTH Aachen, 52074 Aachen, Germany*

e-mail: turchanin@physik.uni-bielefeld.de
Tel.: +49-521-1065376
Fax: +49-521-1066002

**Abstract**

Graphene-based materials have been suggested for applications ranging from nanoelectronics to nanobiotechnology. However, the realization of graphene-based technologies will require large quantities of free-standing two-dimensional (2D) carbon materials with tuneable physical and chemical properties. Bottom-up approaches via molecular self-assembly have great potential to fulfil this demand. Here, we report on the fabrication and characterization of graphene made by electron-radiation induced cross-linking of aromatic self-assembled monolayers (SAMs) and their subsequent annealing. In this process, the SAM is converted into a nanocrystalline graphene sheet with well defined thickness and arbitrary dimensions. Electric transport data demonstrate that this transformation is accompanied by an insulator to metal transition that can be utilized to control electrical properties such as conductivity, electron mobility and ambipolar electric field effect of the fabricated graphene sheets. The suggested route opens broad prospects towards the engineering of free-standing 2D carbon materials with tuneable properties on various solid substrates and on holey substrates as suspended membranes.

**Table of contents graphic**

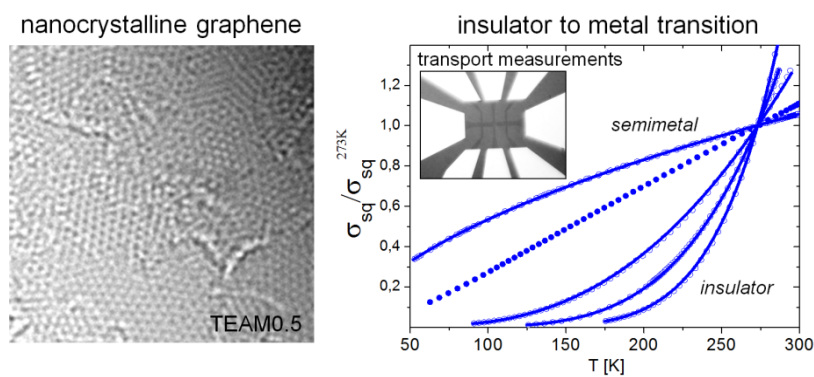

The demonstration of free-standing layers of graphite[1-2] –graphene– has triggered enormous research activity.[3-7] The interest in two-dimensional (2D) carbon materials ranges from the unique electric transport in graphene single crystals[8-9] to the membrane properties of atomically thin free-standing sheets.[10] 2D carbon materials have been suggested for use as transparent conductive coatings,[11] biological filters,[12] molecular sensors,[13] nanoelectromechanical components,[14] transmission electron microscope (TEM) supports,[15-16] or nanocomposite materials,[17-18] just to name a few. However, the development of these applications depends on the availability of 2D carbon materials with tuneable and well-defined physical and chemical properties on various substrates and as suspended membranes. Despite recent progress towards a large scale fabrication of graphene, for example via decomposition of small molecules on metal surfaces,[19-21] growth on SiC wafers[22] and from solid carbon sources[23-24] or by chemical exfoliation of graphite,[17, 25] methods to produce large quantities of 2D carbon materials for various specific applications are still lacking. In this respect, bottom-up approaches via *molecular self-assembly* are very promising.[26-28] By using a manifold of organic molecules with diverse functionalities as the elementary building blocks, self-assembly provides broad prospects towards the engineering of free-standing 2D carbon materials with tuneable properties.

Here we present a bottom-up approach for the fabrication of *nanocrystalline graphene sheets* (single or a few layers) that is based on a sequence of irradiative and thermal treatment of aromatic self-assembled monolayers (SAMs[29]). First, electron irradiation cross-links the SAMs into *~1 nm* thick carbon nanosheets;[30] then vacuum annealing[27, 31] transforms these supramolecular sheets into graphene. We characterized structural properties of the so-prepared graphene with high resolution transmission electron microscopy (HRTEM) and Raman spectroscopy and we measured temperature dependencies of the electrical conductivity. The HRTEM data directly demonstrate the conversion of the supramolecular nanosheets into a covalently bonded network of *graphene nanocrystals*. The corresponding electrical transport data show that this transformation is accompanied by an *insulator to metal transition*. The observed states, *i.e.* insulating or semi-metallic, are determined by the annealing temperature, $T_{an}$, and/or by the thickness of nanosheet stacks. In the presence of a back gate voltage a large ambipolar electric field effect was observed in samples displaying the insulating behaviour. The electrical transport is quantitatively analyzed by a model describing the growth of graphene islands in the nanosheet plane.

**RESULTS**

In Fig. 1 the fabrication route is schematically summarized. First, an aromatic SAM is formed on a substrate, Fig. 1(a). To this end, we used 1,1'-biphenyl-4-thiols (BPT) to form a densely packed SAM on gold with a thickness of *~1 nm*.[32] This SAM is composed of aromatic carbon rings which are the same building blocks as in graphene. Next, the BPT SAM is exposed to electron radiation (electron energy *100 eV*, dose *60 mC/cm$^2$*) in high vacuum, Fig. 1(b). The electrons induce lateral cross-linking of the aromatic rings and transform the monolayer into a supramolecular carbon nanosheet with a thickness of one biphenyl molecule. As the exposure can be performed either by a focused electron beam or by large area flood exposure, sheets with lateral dimensions from *~10 nm* up to several cm and larger can be fabricated.[33] The nanosheet is then annealed in ultra high vacuum (UHV) at temperatures up to *1200 K* to induce its transformation into graphene, *cf.* Fig. 1(c). In a detailed surface analytical study[32] it was found that sulphur initially present in the monolayer, completely desorbs at temperatures above *800 K*. Simultaneously, the carbon rings evolve from a nearly upright orientation, as in a BPT SAM, to an in-plane orientation, as in graphene. Upon increasing the annealing temperature, the transformation towards graphene proceeds, which is confirmed by the appearance of the characteristic G-peak in the Raman spectrum and onset of the electrical conductivity.[27] The thickness of the resulting graphene sheet is given by the thickness of the precursor SAM and is below 1 nm as measured by atomic force microscopy and x-ray photoelectron spectroscopy.[27,32]

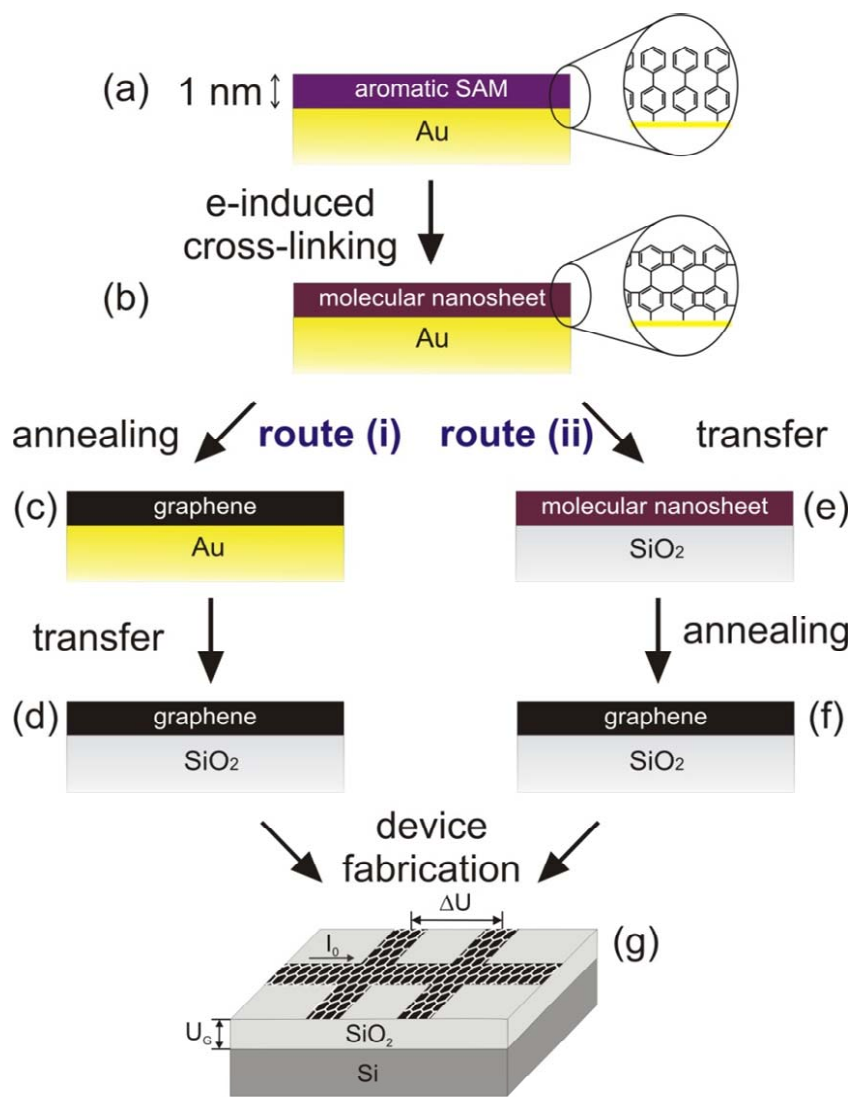

**Figure 1.** Schematic representation of two routes for the fabrication of graphene from aromatic SAMs. (i): (a) Formation of a SAM on a metal substrate (e.g. thiol SAM on Au); (b) electron-induced cross-linking into a supramolecular carbon nanosheet; (c) transformation into graphene via vacuum annealing and (d) subsequent transfer to a new (insulating) substrate. (ii) (a) and (b) are the same as in route (i) followed by transfer (e) to a new substrate and annealing (f). (g) Schematic representation of a Hall bar device with layout for electric measurements.

As shown in Fig 1, we prepared samples by two routes. In route (i) carbon nanosheets were annealed on the gold substrates and then transferred onto oxidized (~300 nm SiO$_2$) silicon (route (i): (a)→(b)→(c)→(d) in Fig.1). To transfer the nanosheets, a polymeric transfer medium (polymethyl methacrylate, PMMA) was spin-coated on the sheet and then the original substrate was dissolved[27, 34] (see experimental methods for details). The PMMA/nanosheet sandwich structure was then placed onto the target surface, and the polymer was dissolved, leaving the nanosheet on a new substrate. This transfer procedure is very efficient and can be also used for a transfer on perforated substrates. In this way free-standing nanomembranes with suspended areas more than $200 \times 200\ \mu m^2$ have been created.[15, 34] Note that in our fabrication procedure, the sequence of transfer and thermal annealing can be exchanged (route (ii): (a)→(b)→(e)→(f) in Fig.1). This is necessary when the initial substrate does not withstand high temperatures or when multiple sheets are stacked onto each other on the target substrate to form a few layers of graphene. For both, routes (i) and (ii), an annealing time of $30\ min$ was applied to allow thermal equilibration.

Structural characterization of the annealed nanosheets was performed by Raman spectroscopy and high resolution transmission electron microscopy (HRTEM). Fig. 2 shows a room temperature (RT) HRTEM image (a-c) and a Raman spectrum (d) of a nanosheet annealed at *1200 K* in UHV. In the latter, the G- and D-peaks at *1605* and *1365 cm$^{-1}$* are identified, that indicates the formation of graphene nanocrystals.[36] The micrograph in Fig 2(a) was obtained with the TEAM0.5, an aberration corrected HRTEM capable of resolving individual atoms of light elements.[37] It is clearly seen, that well ordered graphene areas with lateral dimensions from *2* to *5 nm* have formed after annealing and that they alternate with less ordered areas. These ordered and less ordered domains are covalently bonded into a 2D network consisting of more than 60% of graphene areas. Local Fourier transforms in Fig. 2a confirm a hexagonal symmetry (Fig. 2(c)) of the graphene areas while other areas show a 2D amorphous structure (Fig. 2(b)). These microscopy data demonstrate that the Raman spectrum in Fig. 2(d), which was obtained from an ~*3 μm$^2$* spot, is a superposition of both nanocrystalline and amorphous carbon.

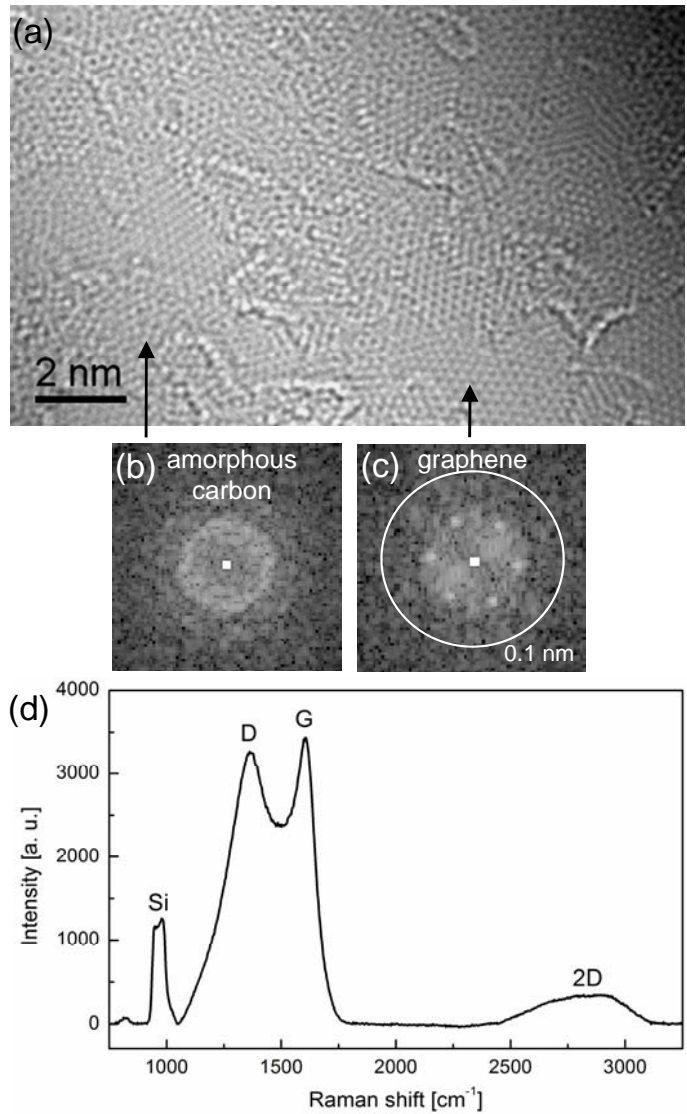

**Figure 2.** (a-c) High-resolution transmission electron microscope image of a fabricated via route (i) graphene, $T_{an} = 1200\ K$. The graphene sheet was transferred onto a TEM grid with the lacey carbon film. The image was recorded at *80 kV* with $C_s = -0.015\ mm$ (3rd-order spherical aberration coefficient[35]) and $C_5 = 5\ mm$ (5th-order spherical aberration coefficient[35]) at an overfocus of ~*9 nm*. (a) Unprocessed image; (b) and (c) local Fourier transforms from *3*x*3 nm²* areas indicated by the arrows. The presence of diffraction spots in (c) verifies the presence of crystalline graphene areas in (a). (d) Raman spectrum (excitation wavelength *532 nm*) of the same sample transferred to a $SiO_2$/Si surface.

Next, we studied temperature dependencies of electrical conductivity, $\sigma(T)$, for samples annealed at different temperatures to characterize the electrical transport. The measurements were conducted by the four-point probe method using Hall bar structures (see Fig. 1(g) and Fig. 3(a, b)). To fabricate them, nanosheets on the oxidized Si wafers were processed by standard micro-fabrication techniques (see experimental methods for details). Fig. 3(a) shows a light microscope (LM) micrograph of a Hall bar structure etched into a nanosheet. The nanosheet areas have a darker contrast in comparison to the bare $SiO_2$. Fig. 3(b) presents the same structure with evaporated gold contacts and a layout for electrical measurements, which were carried out in a temperature range from *50* to *300 K*.

We found that the electrical conductivity of the nanosheets depends strongly on annealing temperature, $T_{an}$, which determines the degree of the transformation into graphene. Fig. 3(c) shows $\sigma(T)$ for monolayer samples prepared by route (i). By varying $T_{an}$ from *900* to *1200 K* the RT conductivity, $\sigma_{sq}^{RT}$, changes five orders of magnitude and approaches a value of *~0.1 mS sq*, which has earlier been measured for single-crystalline graphene sheets obtained by mechanical exfoliation.[8] In general, the conductivity of all samples *increases* with temperature. However, the behaviour of $\sigma(T)$ for samples with high and low conductivity is very different. This difference is clearly seen in Fig. 3(d), where $\sigma(T)$ is plotted after normalization to $\sigma_{sq}^{273\ K}$. Samples with low conductivity (1-3) demonstrate strongly insulating behaviour with a positive curvature of $\sigma(T)$, whereas the sample with the highest conductivity (5) shows a negative curvature of $\sigma(T)$. Sample 4 presents an intermediate case, cf. Fig. 3(c, d). Such a variation of $\sigma(T)$ is a fingerprint of the insulator-metal transition,[38] which evolves in the molecular nanosheets upon their transformation into graphene.

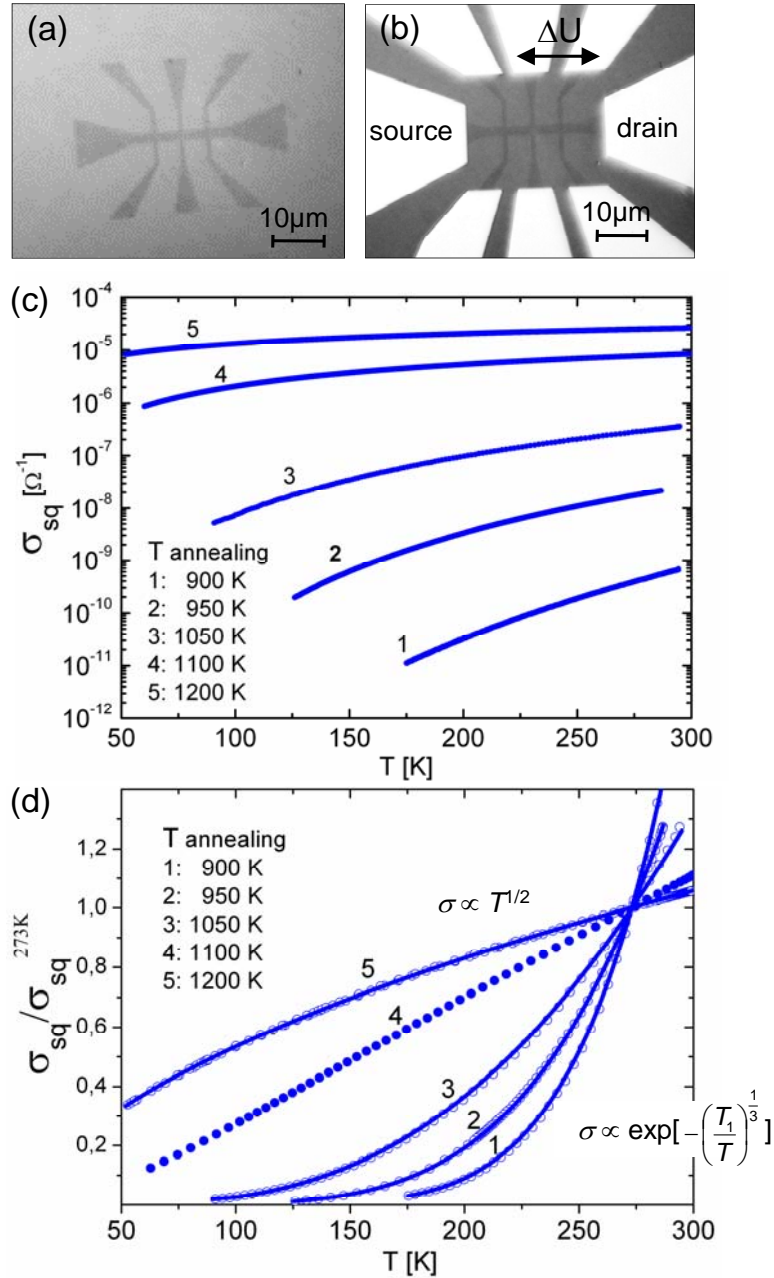

**Figure 3.** Electrical conductivity, $\sigma(T)$, of the monolayer samples annealed at different temperatures. The samples were prepared by route (i). (a) Light microscope images of a Hall bar device fabricated by etching the sheets on a Si wafer with *300 nm* of Si-oxide; (b) the same structure with evaporated gold contacts and a layout for electrical measurements. (c) Experimental $\sigma(T)$ data for the samples annealed at different temperatures. (d) $\sigma(T)$ data presented in (c) after normalization to $\sigma(273\ K)$ of the respective samples (dots). Solid lines are the theoretical curves for the insulating (1-3) and semi-metallic (5) states.

Similar changes in the electrical transport were found for multilayer samples as a function of the number of carbon nanosheets in a stack. After annealing at *1100 K*, the monolayer still does not conduct at RT, whereas two-layer stacks are conducting and thicker stacks show an increase in conductivity, see Fig. 4(a). The delay in onset of the conductivity in comparison to the samples directly annealed on gold may relate to the specific interactions at the $SiO_2$/nanosheet interface, which postpone the graphitization. As seen from Fig. 4(b), the curvature of $\sigma(T)$ changes its sign as the number of nanosheets grows from two to five. Thus, the insulator-metal transition in this two-dimensional carbon system takes place not only by varying the annealing temperature, as shown for monolayers (see Fig.3), but also by changing the number of carbon nanosheets in a stack.

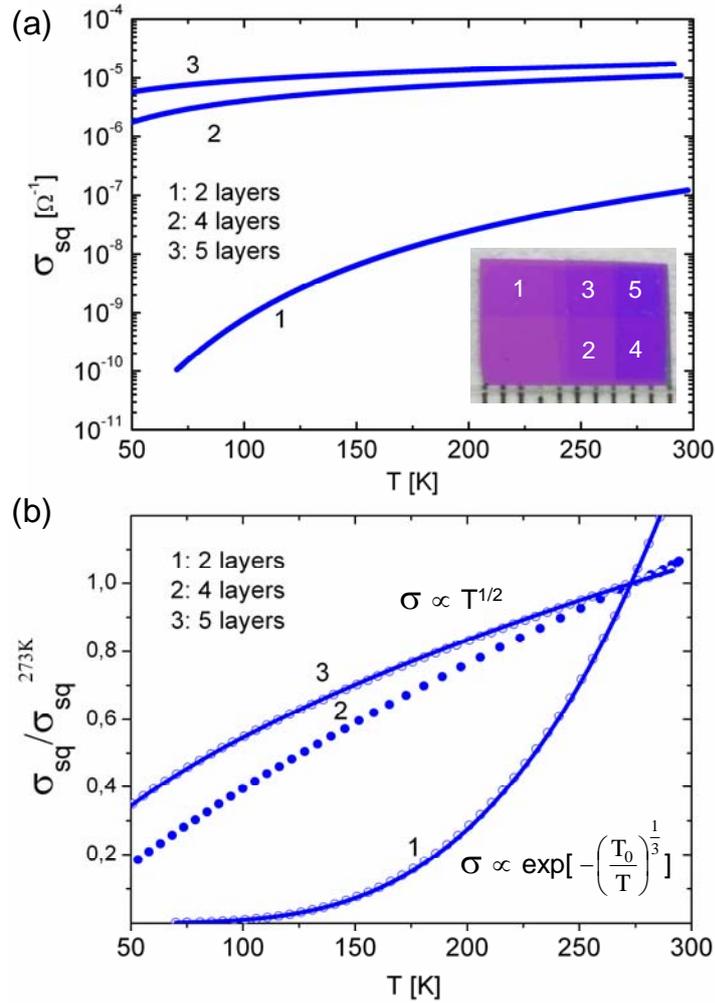

**Fig. 4** Electrical conductivity, σ(T), of multilayer samples after annealing at *1100 K*. The samples (see insert; units in the scale bar correspond to 1 mm) were prepared by route (ii). (a) Experimental σ(T) data as a function of the number of layers in a stack. (b) σ(T) data presented in (a) after normalization to σ(*273 K*) of the respective samples (dots). Solid lines are the theoretical curves for the insulating (1) and semi-metallic (3) states.

**DISCUSSION**

To quantitatively analyze the experimental $\sigma(T)$ data for the monolayer and multilayer samples (Fig. 3 and 4), we propose the following model. HRTEM (Fig. 2(a-c)) shows that after annealing molecular nanosheets transform into a granular 2D system that consists of in-plane oriented nanocrystalline graphene patches alternating with amorphous boundaries. The electrical transport in granular systems crucially depends on the ratio of the tunnelling conductance between neighbouring granules, $\sigma_t$, and the quantum conductance, $\sigma_q = e^2/h$, where $h$ is the Planck constant and $e$ is the electron charge.[38] If this ratio is small, that is $\sigma_t < \sigma_q$, charge carriers (electrons or holes) are strongly localized on the granules and the transport occurs in the form of thermally activated variable range hopping (VRH).[38] In this insulating regime $\sigma(T)$ is governed either by the Efros-Shklovskii or the Mott laws.[39] The Efros-Shklovskii law is realized in systems displaying strong Coulomb interactions between charges on the adjacent granules, and in this case $\sigma(T) \propto \exp\left[-\left(\frac{A}{T}\right)^{\frac{1}{2}}\right]$.[38-41] If the Coulomb interactions are negligible, $\sigma(T)$ is described by the Mott law, which is written for the 2D case as

$$\sigma(T) = \sigma_0 \exp\left[-\left(\frac{T_1}{T}\right)^{\frac{1}{3}}\right] . \qquad (1)$$

The parameter $T_1$ is given by[39]

$$T_1 = \frac{27}{\alpha k_B \rho(E_F) \xi^2} , \qquad (2)$$

where $\rho(E_F)$ is the density of localized states at the Fermi energy $E_F$; $\xi$ is the localization length of the wave functions determined by the tunneling conductance, $\sigma_t$, between neighbouring granules; and the constant $\alpha$ has a value of about 2.[39] In this regime the typical length of the electron hopping, $r$, is given by $r = \frac{\xi}{3}\left(\frac{T_1}{T}\right)^{1/3}$ and the temperature dependent activation energy, $E_0$, is $E_0 = \frac{k_B T}{3}\left(\frac{T_1}{T}\right)^{1/3}$.

By fitting the experimental $\sigma(T)$ data we find that for low transport voltages, $U$, the samples with low conductivity (samples 1-3 in Fig. 3 and sample 1 in Fig. 4) display the 2D Mott law, Eq. (1), in the whole studied temperature range from 90 K to RT. Thus, for samples with a low degree of graphitization the electrical transport is determined by thermally activated VRH and the influence of Coulomb interaction is negligible. Note that similar transport characteristics were previously reported for various amorphous carbon systems[42-43] and graphene oxide.[44-45] However, for samples with high conductivity (sample 5 in Fig. 3 and sample 3 in Fig. 4) in the whole temperature range a power law $\sigma(T) \propto T^p$ with $p \approx 0.5$ is found. This difference is an evidence of the insulator-metal transition taking place in the molecular nanosheets during their transformation into graphene. The non-exponential $\sigma(T)$ dependence strongly indicates an enhanced electrical transport between the neighbouring nanocrystals, that is, a regime where $\sigma_t > \sigma_q$. It was also observed in other semi-metallic systems[46] and can be qualitatively explained by a thermally activated VRH between weakly (non-exponentially) localized electron states.[47-49]

Next, we quantitatively analyse the electric transport data for samples displaying the insulating regime. From the experimental $\sigma(T)$ data (see Fig. 3(b)) and Eq. (1), we extract the parameter $T_1$ for the monolayer samples in the insulating regime:

$T_1(T_{an}=900\ K)=3.3\times10^6\ K$, $T_1(T_{an}=950\ K)=9.5\times10^5\ K$, and $T_1(T_{an}=1050\ K)=2.1\times10^5\ K$. As seen from Eqs. (1) and (2), the transport is determined by two parameters, namely, the localization length, $\xi$, and the density of states, $\rho(E_F)$. We estimate the parameters $\xi$ and $\rho(E_F)$ for the sample annealed at $T_{an}=1050\ K$, because this sample is the closest one to the insulator-metal transition point on the insulator-side. First, to estimate the lower limit of $\xi$ we calculate the hopping length, $r$, and the activation energy, $E_0$, for the lowest temperature ($T=90\ K$), since it gives the most accurate estimation. Thus, we obtain $r=4.4\xi$ and $E_0=3.4\times10^{-2}\ eV$ for $T=90\ K$. Since in the studied temperature range the Coulomb interactions can be neglected, the Coulomb energy, $e^2/(4\pi\varepsilon_0\varepsilon r)$, is smaller than $E_0$, that gives $\xi \geq e^2/(17.6\pi\varepsilon_0\varepsilon E_0)$ or numerically $\xi \geq 2.7\ nm$. Here a value of the dielectric constant $\varepsilon=3$, experimentally measured in thin amorphous carbon films with incorporated aromatic rings,[50] was used.

To estimate the upper limit of $\xi$ we measured current-voltage characteristics, $I(U)$, at high electric fields and at varying temperatures. As presented in Fig. 5(a), the observed non-linearities in $I(U)$ (dots) can be well described (solid lines) by equation:

$$\sigma(T,U) = \sigma(T,0)\exp\left(\frac{\gamma\sqrt{|U|}}{T}\right) \quad (3)$$

where $\gamma$ is a constant. Since the $I(U)$ data show some asymmetry, different values of $\gamma$ for positive and negative voltages were employed. Note that the $I(U)$ data were acquired by the four-probe method (*cf.* Fig. 3(b)), thus the influence of a contact resistance on the observed non-linearities and asymmetry can be excluded. Similar asymmetries in $I(U)$ were also reported for reduced graphene oxide monolayers,[45] Their physical reasons have to be further investigated.

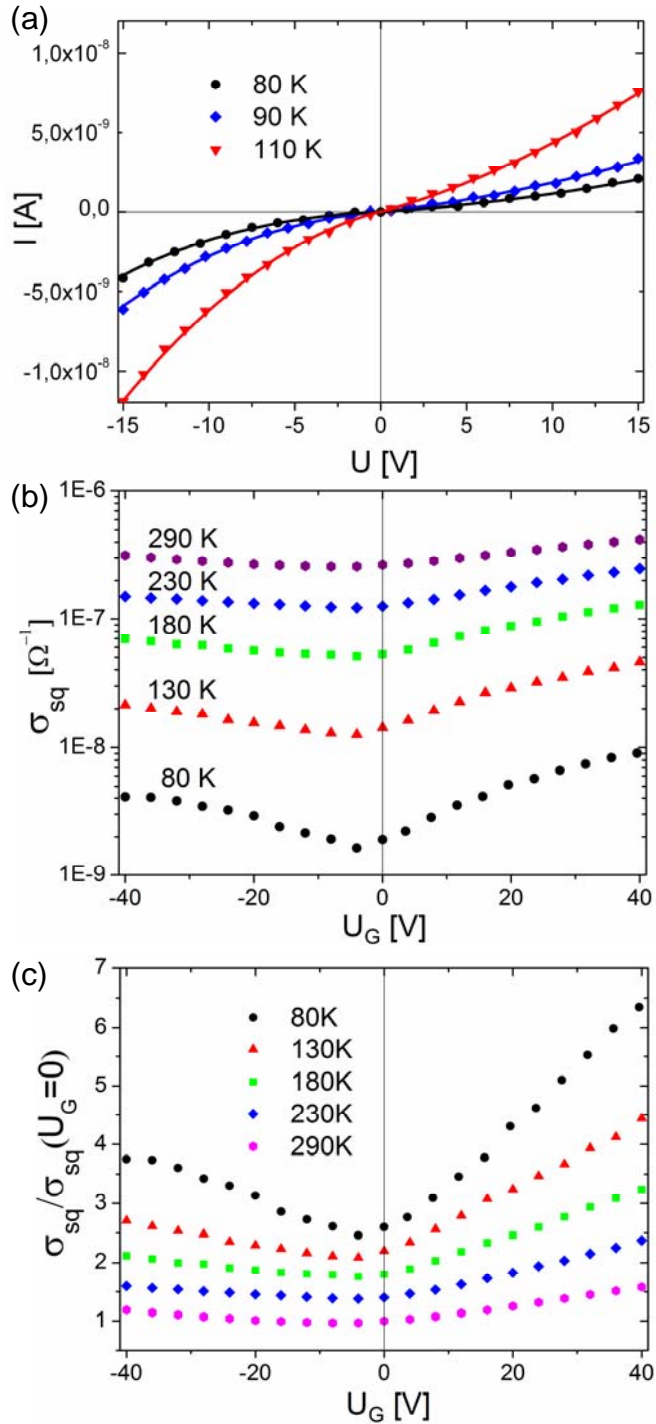

**Fig. 5** Electric characteristics of a monolayer sample prepared via route (i) after annealing at *1050 K* (sample 3 in Fig. 3(c)). (a) Temperature dependent *I(U)* characteristics at high electric fields. Dots are experimental data and solid lines are fits with Eq. (3). (b) Ambipolar electric field effect of the electrical conductivity, $\sigma(U_G)$, measured at a source-drain voltage of *2 V* as a function of temperature. (c) $\sigma(U_G)$ data after normalization to $\sigma(U_G=0)$. For the presentation clarity curves in (c) are shifted by *0.5* with respect to each other along the ordinate axis.

Strong temperature dependencies of the electrical conductivity in high electric fields, as presented by Eq. (3), have been earlier reported for various carbonaceous systems[51-52] demonstrating the VRH transport. They imply that in this transport regime the VRH activation energy, $E_0$, is higher than the energy induced by an external electric field on a distance equal to the hopping length, $r$. This field-induced energy is given by $eUr/L$, where $U$ is a drop of voltage on the sample length $L$. Thus, for the $I(U)$ data presented in Fig. 5(a), the inequality $eUr/L \leq E_0$ holds in the whole range of applied voltages. With this inequality and using the earlier obtained expression of $r=4.4\xi$ and $E_0$ for $T=90\ K$ and the maximum value of $U$ (15 V), the upper limit of the localization length can be estimated as $\xi \lesssim LE_0/(4.4eU_{max}) \sim 7\ nm$. Combining the upper limit with the lower limit, estimated before, the localization length is given by $2.7\ nm \leq \xi \lesssim 7nm$. Note, if $E_0$ is smaller in comparison to the energy induced by an external electric field, i.e. $eUr/L > E_0$, the transport is determined by the non-activated electric field induced hopping.[38-41] In this case the conductivity in high electric fields strongly depends on voltage and no temperature dependence as described by Eq. (3) is observed.

Since an insulator-metal transition usually takes place as the distance between conducting nanocrystals, $d_{am}$, becomes of the order of $\xi$,[36] we estimate the size of amorphous boundaries between graphene nanocrystals near the transition point as $d_{am} \approx \xi \sim 5\ nm$. Thus, for samples displaying an insulating behaviour, the size of graphene nanocrystals corresponds to $\xi < d_{am}$ and is less than $5\ nm$. However, graphene areas grow with increasing $T_{an}$ and amorphous areas shrink, which implies that $d_{am} < \xi$ in a semi-metallic regime, and it also gives the size of graphene nanocrystals on the metal-side of the transition to be ~5 nm. This value is in very good agreement with microscopy data, Fig. 2(a), and it shows that the electric transport properties can be well described by the applied model.

As seen from Eqs. (1)-(2), the conductivity of samples in the VRH regime strongly depends on $\rho(E_F)$. Since by applying a gate voltage (*cf.* Fig. 1(g)) the Fermi energy of a sample is shifted, the density of states corresponding to the new Fermi energy level can be different, $\rho(E_F(U_G))$. A small modification of the density of states will lead to a large ambipolar electric field effect in the electrical conductivity. An experimental observation of the electric field effect for a sample showing the VRH regime (sample 3 from Fig. 3) as a function of the applied back gate voltage in the range from -40 V to 40 V is presented in Fig. 5(b, c). Its variation on positive and negative values of $U_G$, cf. Fig. 5(c), indicates a two-band structure of the density of states. Assuming a weak linear dependence[53] for $\rho(E_F(U_G))$, the quantitative analysis can be performed using the following expression:

$$\rho(E_F(U_G)) = \rho_0(1 + \beta_{e(h)} |E_F(U_G)|), \qquad (4)$$

where $\rho_0$ is the density of states at $E=0$. Substituting $\rho(E_F)$ in Eq. (2) by $\rho(E_F(U_G))$ we obtain with the help of Eq. (1) a relation between the conductivity and gate voltage in the following form:

$$\ln\left[\frac{\sigma(U_g)}{\sigma(0)}\right] = \frac{1}{3}\left(\frac{T_1}{T}\right)^{1/3} \beta_{e(h)} (|E_F(U_G)| - |E_F|). \qquad (5)$$

The shift of the Fermi energy in Eq. (5) is given as $\dfrac{\varepsilon_{gi}\varepsilon_0 U_G}{eD} = \int\limits_{E_F}^{E_F(U_G)} dE\rho(E)$, where $D$ and $\varepsilon_{gi}$ are a thickness and a dielectric constant of the gate insulator (SiO$_2$, $D=300$ nm, $\varepsilon_{gi}=3.7$), respectively. Assuming a small variation of the density of states, this expression can be presented in a simplified form as:

$$E_F(U_G) = E_F + \frac{\varepsilon_{gi}\varepsilon_0 U_G}{eD\rho(E_F)} \quad . \tag{6}$$

As seen from Eqs. (5)-(6), the gate effect becomes more pronounced at low temperatures and diminishes towards RT, which is in very good agreement with the experimental results, *cf.* Fig. 5(b, c). Using Eq. (5) we extract a magnitude of the Fermi energy, which is ~*0.1 eV*. Note that the linear dependence of $\rho(E)$ in graphene nanocrystals is much weaker in comparison to the single crystal graphene and shows an essential electron/hole asymmetry with $\beta_e$~*0.5 (eV)$^{-1}$* > $\beta_h$~*0.3(eV)$^{-1}$*.

The field effect mobility, $\mu$, can be obtained from the linear regions of the $\sigma(U_G)$ data presented in Fig. 5 (b, c) assuming that:

$$\sigma(U_G) = \sigma_{min}(U_G^*) + \mu \frac{\varepsilon_{gi}\varepsilon_0(U_G - U_G^*)}{D} \quad , \tag{7}$$

where $\sigma_{min}(U_G^*)$ is the minimum conductivity at which the charge neutrality condition is reached and $U_G^*$ is the respective gate voltage. A slight shift of $U_G^*$ (less than by 5 V) towards negative voltages was observed, *cf.* Fig. 5 (b, c). Thus, for *T=80 K* the mobilities of electrons and holes of ~*0.02 cm$^2$/Vs* and ~*0.01 cm$^2$/Vs* were found, respectively. These values linearly increase with temperature reaching the values of ~*0.5 cm$^2$/Vs* and ~*0.2 cm$^2$/Vs* at RT. The concentration of charge caries at RT near $U_G$=0 V can be estimated using the general expression for conductivity as $n_0=\sigma/(e\mu)$, which gives the value of ~*4×10$^{12}$ cm$^{-2}$*. Since this value is obtained for a sample near the insulator-metal transition point and since on the metal side a strong increase of $n_0$ is not expected[39] (which is opposite on the insulator side), we can roughly estimate the RT mobility for sample 5, that is, for a sample in the semi-metallic regime. Using the obtained $n_0$ value, the maximum electron mobility is found to be ~*40 cm$^2$/Vs*.

Although this value can be somewhat overestimated, it reasonably correlates with microscopy data and model description. It shows an increase of the electrical transport with shrinking of amorphous grain boundaries and growth of graphene crystallites in a covalently bonded 2D carbon network after annealing of nanosheets at high temperatures.

**CONCLUSIONS**

We have demonstrated a bottom-up route for the large area fabrication of covalently bonded nanocrystalline graphene sheets from self-assembled monolayers. Electron irradiation transforms aromatic SAMs into molecular nanosheets with a thickness of *~1 nm*. Subsequent vacuum annealing induces the formation of graphene nanocrystals oriented in the nanosheet plane. Electrical transport data show that this structural transformation is accompanied by an insulator to metal transition as the graphene areas grow with annealing temperature. The suggested route opens up broad prospects to the fabrication of graphene layers with well controlled thickness and tunable electrical properties on various metal and insulator as well as on solid and holey substrates. The choice of the molecular precursors will facilitate the optimization of the properties for specific applications including the crystallinity. Note that the demonstrated covalently bonded networks of graphene nanocrystals may be of interest for the investigation of magnetism in 2D carbon systems.[54]

# EXPERIMENTAL METHODS

## Fabrication and transfer of graphene sheets

First, 1,1'-biphenyl-4-thiol (BPT) SAMs were prepared on *300 nm* thermally evaporated gold on mica substrates (Georg Albert PVD-Coatings). The substrates were cleaned in a UV/ozone-cleaner (FHR), rinsed with ethanol and blown dry in a stream of nitrogen. They were then immersed in a *~1 mmol* solution of BPT in dry, degassed dimethylformamide (DMF) for *72 h* in a sealed flask under nitrogen. Afterwards the samples were rinsed with DMF and ethanol and blown dry with nitrogen. The crosslinking (transformation into molecular nanosheets) was achieved in high vacuum ($<5*10^{-7}$ *mbar*) with an electron flood gun (Specs) at an electron energy of *100 eV* and a dose of *60 mC/cm$^2$*.

To induce the transformation of nanosheets into graphene they were annealed on gold surfaces or silicon wafers with a *300 nm* layer of silicon oxide (after transfer) at UHV conditions in Mo sample holders with a resistive heater. The heating/cooling rates of *~150 K/h* and the annealing time of *0.5 h* were applied. Annealing temperature was controlled with a Ni/Ni-Cr thermocouple and a two-color pyrometer (SensorTherm). The nanosheets on gold were annealed in vacuum at temperatures up to *~1200 K*. Since the mica substrate is damaged above *~1000 K*, for annealing at temperatures above *1000 K*, the gold films with the nanosheets on top were cleaved from the mica by immersion in hydrofluoric acid (*48%*) and transferred onto clean quartz substrates.

To transfer the non-annealed and annealed nanosheets to a new substrate a *~500 nm* thick layer of polymethylmethacrylate (PMMA) was spin-coated and baked on their surfaces.[15] This layer was used for mechanical stabilization of the nanosheets during transfer. Then, the gold was cleaved from the mica by immersion in hydrofluoric acid and etched away in an $I_2$/KI-etch bath. Afterwards the nanosheet/PMMA was transferred onto a $SiO_2$ substrate or TEM grid and the PMMA was dissolved in acetone to yield a clean nanosheet. The cleanness of the nanosheets after transfer was confirmed by X-ray photoelectron spectroscopy.[12, 27] By repeating the transfer of nanosheets on top of each other the multilayer samples were prepared. In this way large scale free-standing nanosheet/graphene layers can be prepared on any surface or as suspended membranes.

**Spectroscopy and Microscopy**

Raman spectra were collected at the excitation wavelength of *532 nm* using a micro Raman spectrometer LabRam ARAMIS equipped with a frequency-doubled Nd:YAG-Laser and HeNe Laser, a *100*x objective and a thermoelectrically cooled CCD detector ( *2-3 $cm^{-1}$* spectral resolution). The Si-peak at *520.5 $cm^{-1}$* was used for peak shift calibration of the instrument.

Electron microscopy was performed with TEAM0.5.[35] Briefly, the microscope is equipped with a high brightness gun, a monochromator, a pre- and a post-specimen aberration corrector, and an ultra twin lens. Electrical and mechanical stabilities allow for *0.05 nm* resolution at *300 kV* and sub Ångstrom resolution at *80 KV* (*<0.08 nm*). Images were recorded with the monochromator excited to provide an energy spread of *0.15 eV*.

**Microfabrication of Hall bar devices**

The Hall bar devices were fabricated from the pyrolized nanosheets which were placed on silicon wafers with a *300 nm* thick gate oxide layer. The fabrication process included three lithographic steps. Step 1: Spin-coating of a resist layer (PMMA) on the surface; pattern transfer into the layer by electron beam lithography (EBL) (Vistec

EBPG 5000+) and resist development; reactive ion etching in an oxygen/argon plasma (Leybold Z401) of the non-protected areas of nanosheets; dissolution of PMMA in acetone. This step results in the fabrication of a device structure on a silicon oxide surface as shown in Fig. 3(a) of the manuscript. Step 2: Spin-coating of a new PMMA layer; electron beam lithography and development of the resist layer; subsequent vacuum evaporation ($p < 10^{-6}$ *mbar*) of a *10 nm* adhesive Ti layer and a *40 nm* Au layer; lift-off of the resist. This procedure leads to the manufacturing of metal contacts to the device bars as shown in Fig. 3(b) of the manuscript. Step 3: We have found out that the quality of the metal/nanosheet contacts could be improved significantly if an additional *100 nm* thick Au layer, having a direct contact to the device bars, was evaporated. This additional gold layer was fabricated by a similar procedure as described in Step 2.

After Steps 1-3 a wafer with the manufactured Hall bar devices was fixed onto a chip carrier with conductive silver epoxy glue. Then gold wires were bonded (Westbond 5700) to the device to from electrical contacts with the chip carrier pins. By bonding special care was taken not to damage the underlying back gate oxide layer.

**Electrical measurements**

For characterization of the Hall bar devices a B1500A Semiconductor Parameter Analyzer (Agilent Technologies) was employed. This instrument is equipped with seven source-measure units which can be individually connected to the source, drain and side contacts of a Hall bar device (see Fig. 3(b)). Such a measuring arrangement was found to be particularly useful for the characterization of samples with high resistances (more than *1 MΩ*), as it allows to avoid parasitic leakage currents through the measurement setup. The 2D conductivity, $\sigma_{sq}$, of the samples was measured by four-point method and derived from the source-drain current, $I_0$, the drop of voltage at the respective side contacts of a Hall bar device, $\Delta U$, the distance between the contacts, $l$, and the width of the Hall bar, $w$, as given by

$$\sigma_{sq} = \frac{I_0}{\Delta U} \frac{l}{w} \ [\Omega^{-1}\text{sq}].$$

The typical width of the manufactured Hall bar devices was *1 µm* and the length between source and drain was *15 µm*, the distance between side contacts was *10 µm*. To find $\sigma_{sq}$ values linear *I(U)* characteristics were measured. At these measurements the drop of voltage between source and drain was typically of *~1 V*. The temperature dependence of electrical conductivity, $\sigma_{sq}(T)$, was studied in the range from *50* to *300 K* in a cryostat (Oxford Optistat).

**Acknowledgment**

We thank C.T. Nottbohm (University of Bielefeld) and S. Wundrack (PTB) for technical assistance. Financial support from the Volkswagenstiftung, Deutsche Forschungsgemeinschaft (SPP "Graphene"), and the Germany Ministry for Education and Research (BMBF) and participation in the COST Action CM0601 (Electron Controlled Chemical Lithography, ECCL) are acknowledged.